\documentclass[aps,prc,twocolumn,preprintnumbers,amsmath,showpacs,floatfix,nofootinbib]{revtex4-1}
\usepackage{graphicx}
\usepackage{bm}
\usepackage{color}

\newcommand{\dd}{{\rm d}}
\newcommand{\ee}{{\rm e}}
\newcommand{\ii}{{\rm i}}
\newcommand{\mean}[1]{\left\langle #1 \right\rangle} 

\begin{document}


\author{Nicolas Borghini} \email{borghini@physik.uni-bielefeld.de}
\author{Steffen Feld} \email{s.feld@physik.uni-bielefeld.de}
\author{Nina Kersting} \email{nkersting@physik.uni-bielefeld.de}
\affiliation{Fakult\"at f\"ur Physik, Universit\"at Bielefeld, 
	Postfach 100131, D-33501 Bielefeld, Germany}

\title{Scaling behavior of anisotropic flow harmonics in the far from equilibrium regime}

\date{\today}

\begin{abstract}
  We consider the development of anisotropic flow in an expanding system of particles undergoing very few rescatterings, using a kinetic-theoretical description with a nonlinear collision term. 
  We derive the scaling behaviors of the harmonic coefficients $v_n$ with the initial-state eccentricities and the mean number of rescatterings, and argue that hexagonal flow $v_6$ should follow a nontrivial behavior, different from that of the lower harmonics.  
  Our findings 
  should be observable in experimental data for small systems.
\end{abstract}


\maketitle

\section{Introduction}
\label{s:intro}

A highlight of the results from the ongoing experimental programs with heavy nuclei at the CERN LHC or the Brookhaven RHIC consists of the measurements of various azimuthal correlations between outgoing particles. 
In particular, the measured values of Fourier coefficients $v_n$ that quantify the anisotropies of the transverse emission pattern are interpreted as footprints of a strongly collective behavior~\cite{Heinz:2013th}, hinting at the creation of a medium close to local thermodynamic equilibrium.

According to the most widely accepted theoretical picture, the final-state anisotropies in momentum space reflect asymmetries of the spatial geometry of the ``initial state''---i.e.\ of the distribution of entropy density in the transverse plane after the nuclei have passed through each other~\cite{Ollitrault:1992bk,Alver:2010gr}---, characterized for instance by ``eccentricities''~\cite{Teaney:2010vd,Gardim:2011xv}
\begin{equation}\label{eccentricities}
\epsilon_n\ee^{\ii n\Phi_n} \equiv -\frac{\mean{r^{n\,}\ee^{\ii n\theta}}}{\mean{r^n}}
\end{equation}
for $n\geq 2$, and by generalizations thereof.\footnote{Throughout the paper we ignore for simplicity the first harmonic $v_1$ and the corresponding eccentricity $\epsilon_1$, whose standard definition differs from Eq.~\eqref{eccentricities}.}
In this definition, $r$ and $\theta$ are centered polar coordinates in the transverse plane, while the angular brackets denote an average over the entropy density. 
To be more specific, let us briefly quote a number of model findings, without any attempt at exhaustiveness (see also Ref.~\cite{Liu:2018hjh} for a compilation of experimental results from the LHC):
\begin{itemize}
\item To a good approximation, the (integrated) elliptic flow $v_2$ and triangular flow $v_3$ scale linearly with the corresponding eccentricities~\cite{Ollitrault:1992bk,Alver:2010gr,Teaney:2010vd,Gardim:2011xv,Niemi:2012aj,Plumari:2015cfa}:
  \begin{equation}\label{v2,v3_vs_eps2,eps3}
  v_2 \simeq {\cal K}_{2,2}\epsilon_2\quad,\quad
  v_3 \simeq {\cal K}_{3,3}\epsilon_3.
  \end{equation}
  In collisions with a large impact parameter, i.e.\ a larger $\epsilon_2$, a cubic deviation to this behavior for $v_2$ has been reported~\cite{Noronha-Hostler:2015dbi}.
  
\item The quadrangular flow $v_4$ receives two kinds of contributions: a linear scaling with $\epsilon_4$ and a quadratic dependence on $\epsilon_2$~\cite{Borghini:2005kd,Gardim:2011xv,Teaney:2012ke,Niemi:2012aj,Lang:2013oba}:
  \begin{equation}\label{v4_vs_eps2,4}
  v_4 \simeq {\cal K}_{4,4}\epsilon_4 + 
    {\cal K}_{4,22}\epsilon_2^2.
  \end{equation}
  Similarly, the pentagonal flow $v_5$ depends linearly on $\epsilon_5$ and nonlinearly on the second- and third-harmonic eccentricities~\cite{Gardim:2011xv,Teaney:2012ke,Lang:2013oba}:%
  \begin{equation}\label{v5_vs_eps2,3,5}
  v_5 \simeq {\cal K}_{5,5}\epsilon_5 + 
    {\cal K}_{5,23}\epsilon_2\epsilon_3.
  \end{equation}
  In both cases, the linear term is only visible in ultra-central collisions~\cite{Plumari:2015cfa}, while the nonlinear contribution dominates at larger impact parameters.
  
\item Starting with hexagonal flow $v_6$, there appear more than one nonlinear terms at the ``leading order'' that mostly contributes in noncentral collisions~\cite{Bravina:2013ora,Qian:2016fpi,Giacalone:2018wpp}:
  \begin{equation}\label{v6_vs_eps2,3,6}
  v_6 \simeq {\cal K}_{6,6}\epsilon_6 + 
    {\cal K}_{6,33}\epsilon_3^2 + 
    {\cal K}_{6,24}\epsilon_2\epsilon_4 + 
    {\cal K}_{6,222}\epsilon_2^3.
  \end{equation}
  Note already that there are both quadratic and cubic contributions to $v_6$, which will be of relevance for one of the findings of this paper. 
\end{itemize}

The leading candidate theory for describing the evolution of the medium created in collisions of heavy nuclei at ultrarelativistic energies is nowadays relativistic fluid dynamics, and accordingly most of the studies quoted above were performed within that framework.
The only exception is Ref.~\cite{Plumari:2015cfa}, which relies on a kinetic transport approach, however pushed into a regime where it ``mimics'' dissipative fluid dynamics.
In the corresponding language, one finds that the linear response coefficients ${\cal K}_{n,n}$ depend on the transport properties---mostly, the shear and bulk viscosity---of the expanding medium. 
This also holds for the nonlinear response coefficients ${\cal K}_{4,22}$, ${\cal K}_{5,23}$, ${\cal K}_{6,222}$, ${\cal K}_{6,33}$\dots, yet it was argued in Ref.~\cite{Teaney:2012ke} that they are less damped by viscous effects than the linear coefficients. 

The collective-behavior picture underlying the fluid-dynamical interpretation of the anisotropic flow data has however been challenged by the observations in the past few years of similar signals in collisions of ``smaller systems'' like proton-lead, deuteron-gold or even proton-proton, in events with a relatively large number of particles in the final state (see Ref.~\cite{Nagle:2018nvi} for a recent review).

Accordingly, there has been a revival of models with ``weak final-state collectivity'' aiming at investigating generic behaviors of the anisotropic flow coefficients $v_n$ in a regime where the outgoing particles undergo in average very few rescatterings~\cite{Heiselberg:1998es,Borghini:2010hy,He:2015hfa,Romatschke:2018wgi,Kurkela:2018ygx}.
The present paper, which will remain at a semi-qualitative yet general level, is a further step in that direction, identifying a scaling behavior of the higher harmonics, in particular of the hexagonal flow $v_6$, which to our knowledge has not been noted before.%
\footnote{Detailed calculations within a specific setup will be reported in a forthcoming paper~\cite{NB-NK_inprep}.}

\section{Anisotropic flow far from equilibrium}
\label{s:main-section}

Since our goal is to model a situation in which the final-state particles have only rescattered very little, a natural framework is to treat them as particles all along the system evolution---they never form a continuous medium---and to resort to a kinetic theory framework, as we now detail. 

Without loss of generality for our argumentation, we consider a single particle type and introduce its on-shell phase space distribution $f(t,{\bf x},{\bf p})$, where three-vectors are denoted in boldface. 
The evolution of this distribution will generically be described by a kinetic equation of the form~\cite{DeGroot:1980dk} 
\begin{equation}\label{Boltzmann-eq}
p^\mu\partial_\mu f(t,{\bf x},{\bf p}) = -{\cal C}[f],
\end{equation}
with ${\cal C}[f]$ the collision term modeling the effect of rescatterings.\footnote{We use a metric with positive signature $(-,+,+,+)$.} 
As initial condition for the evolution, we consider the following phase space distribution at some time $t_0$, in which the dependences on the space and momentum variables factorize:
\begin{align}
f(t_0,{\bf x},{\bf p}) \sim F({\bf p})_{}G(r)\bigg[1   
&+ \tilde{\epsilon}_2\bigg(\!\frac{r}{R}\!\bigg)^{\!\!2}\cos[2(\theta\!-\!\Phi_2)] \cr
& + \tilde{\epsilon}_3\bigg(\!\frac{r}{R}\!\bigg)^{\!\!3}\cos[3(\theta\!-\!\Phi_3)] + \cdots\bigg], \cr
\label{f(t0,x,p)}
\end{align}
where we again use polar coordinates $(r,\theta)$ for the projection of ${\bf x}$ onto the transverse plane, and discard the dependence on the longitudinal coordinate, which is irrelevant in the following. 
As hinted at by the notations, $F({\bf p})$ is a position-independent momentum distribution, $G(r)$ depends only on the radial coordinate, while $R$ is a length scale ensuring that $\tilde{\epsilon}_2$ and $\tilde{\epsilon}_3$ are dimensionless. 
As noted by Teaney and Yan~\cite{Teaney:2010vd}, the successive powers of $r$ in the square brackets should actually be regulated by some cutoff function to ensure that the distribution $f$ remains positive, yet we did not denote that regulator since it can remain unspecified for our purposes.
Obviously, computing the eccentricities~\eqref{eccentricities} with distribution~\eqref{f(t0,x,p)} (instead of the corresponding entropy density, yet at the present stage this is only a matter of convention) gives 
$\epsilon_n\propto\tilde{\epsilon}_n$ for every $n\geq 2$.

Letting now the initial distribution~\eqref{f(t0,x,p)} evolve, we first consider the collisionless case, ${\cal C}[f]=0$.
The corresponding solutions of the equation of motion~\eqref{Boltzmann-eq} are free-streaming solutions
\begin{equation}\label{free-streaming_sol}
f^{(0)}(t,{\bf x},{\bf p}) = f^{(0)}(t_0,{\bf x}\!-\!{\bf v}(t\!-\!t_0),{\bf p})
\end{equation}
with ${\bf v}$ the velocity corresponding to momentum ${\bf p}$. 
The resulting anisotropic flow coefficients $v_n$,\footnote{For the sake of brevity, we do not write the dependence of $v_n$ on the (modulus of) transverse momentum $p_t$ and on longitudinal momentum / rapidity.} whose calculation involve integrations over the whole position space as well as over the momentum azimuthal angle $\phi_{\bf p}$, are easily shown to be entirely determined by the initial transverse anisotropies of $F({\bf p})$, and in particular independent of the eccentricities $\epsilon_n$. 
Thus, if there is no anisotropic flow initially, as we shall from now on assume, there is none in the final state if the particles do not rescatter, as has long been known. 

To turn on a small amount of rescatterings, we apply the idea whose various incarnations in the heavy-ion physics community went through the successive appellations ``low density limit''~\cite{Heiselberg:1998es,Kolb:2000fha}, ``far from equilibrium regime''~\cite{Borghini:2010hy} and more recently ``eremitic expansion''~\cite{Romatschke:2018wgi} or ``one-hit dynamics''~\cite{Kurkela:2018ygx}, and write a solution of the kinetic equation~\eqref{Boltzmann-eq} with collision term as
\begin{equation}\label{eremitic-Ansatz}
f(t,{\bf x},{\bf p}) = f^{(0)}(t,{\bf x},{\bf p}) + f^{(1)}(t,{\bf x},{\bf p})
\end{equation}
where $f^{(0)}$ is the free-streaming solution~\eqref{free-streaming_sol} and $f^{(1)}$ a small correction term. 
Inserting this ansatz in Eq.~\eqref{Boltzmann-eq} yields at once
\begin{equation}\label{Boltzmann-eq_with_eremitic-Ansatz}
p^\mu\partial_\mu f^{(1)}(t,{\bf x},{\bf p}) = -{\cal C}\big[f^{(0)}\!+\!f^{(1)}\big].
\end{equation}
Multiplied by $\cos(n\phi_{\bf p})$ and integrated over ${\bf x}$ and $\phi_{\bf p}$, the term on the left hand side of this equation yields (up to a normalization factor) the negative of the time derivative $\partial_t v_n$~\cite{Borghini:2010hy,NB-NK_inprep}. 
Integrating over time will then give the final $v_n$.
That is, specific integrals of the collision term ${\cal C}\big[f^{(0)}\!+\!f^{(1)}\big]$ yield the $n$-th anisotropic flow harmonic $v_n$. 

Since we are interested in the development of the higher harmonics, and in particular in the nonlinear contributions, a natural choice for the collision term is Boltzmann's collision integral for elastic two-to-two scatterings, which we symbolically write in the form
\begin{align}\label{collision-integral_2->2}
{\cal C}[f({\bf 1})] = \int_{{\bf p_2},{\bf p_3},{\bf p_4}}\!\big[&
  f({\bf 3})f({\bf 4})_{}w({\bf 3}\!+\!{\bf 4}\to{\bf 1}\!+\!{\bf 2}) \cr
   &- f({\bf 1})f({\bf 2})_{}w({\bf 1}\!+\!{\bf 2}\to{\bf 3}\!+\!{\bf 4})\big],\qquad
\end{align}
where the shorthand notation $f(\bm{j})$ stands for $f(t,{\bf x},{\bf p}_j)$ while the terms $w(\bm{i}+\bm{j}\to \bm{k}+\bm{l})$ involve transition probabilities and the necessary $\delta$-distributions implementing energy-momentum conservation.
Note that in contrast to Ref.~\cite{Borghini:2010hy} we need not assume that the initial geometry is invariant under the ${\bf x}\to -{\bf x}$ transformation, nor that the involved interactions are parity non-violating.
Given a model for the interaction, $w$ will be proportional to some typical cross section $\sigma$. 
In turn, if one computes the total number of rescatterings taking place over the whole system evolution, it will also be approximately proportional to $\sigma$, as will be the average number of rescatterings per particle $\bar{N}_{\rm resc.}$.
The latter constitutes the dimensionless parameter that quantifies the smallness of $f^{(1)}$ relative to $f^{(0)}$.

When substituting $f$ by $f^{(0)}+f^{(1)}$ in this collision integral, we make use of the fact that $f^{(1)}$ is assumed to be a small correction and approximate~\cite{Heiselberg:1998es,Borghini:2010hy,NB-NK_inprep}
\begin{equation}\label{approximate_collision-term}
{\cal C}\big[f^{(0)}\!+\!f^{(1)}\big] \simeq {\cal C}\big[f^{(0)}\big].
\end{equation}
That is, the free-streaming solution~\eqref{free-streaming_sol} fully determines the collision term and thereby the flow coefficients. 

Performing the necessary calculations requires specific models for the as yet unspecified functions $F({\bf p})$ and $G(r)$ in the initial-state distribution~\eqref{f(t0,x,p)} and for the interaction. 
General scalings can however already be predicted irrespective of any specific choice, which we now list. 
\begin{itemize}
\item The multiplication of the isotropic term in one of the factors $f^{(0)}(\bm{i})$ in the integrand of the collision integral~\eqref{collision-integral_2->2} with the term in $\bar{\epsilon}_n\cos[n(\theta-\Phi_n)]$ in the associated $f^{(0)}(\bm{j})$ yields a contribution to $v_n$ proportional to $\sigma_{}\epsilon_n$, i.e.\ approximately proportional to $\bar{N}_{\rm resc.}\epsilon_n$. 
With the values of the eccentricities relevant for heavy ion collisions, this is the dominant contribution to $v_2$ and $v_3$, resulting in linear scalings of the form
  \begin{equation}\label{v2,v3_vs_eps2,eps3,sigma}
  v_2 \sim \bar{N}_{\rm resc.}\kappa_{2,2\,}\epsilon_2\quad,\quad
  v_3 \sim \bar{N}_{\rm resc.}\kappa_{3,3\,}\epsilon_3.
  \end{equation}
For $n\geq 4$, other contributions to $v_n$ are likely to be as important, which we now discuss.
  
\item Besides the linear term in $\sigma_{}\epsilon_4$, another contribution to $v_4$ is generated by multiplying together the terms in $\epsilon_2\cos[2(\theta-\Phi_2)]$ in both distributions $f^{(0)}(\bm{i})$, $f^{(0)}(\bm{j})$ of one of the products $f^{(0)}(\bm{i})f^{(0)}(\bm{j})$ in the integrand of Eq.~\eqref{collision-integral_2->2}. 
  Thus, one obtains
  \begin{equation}\label{v4_vs_eps2,sigma}
  v_4 \sim \bar{N}_{\rm resc.}\kappa_{4,4\,}\epsilon_4 + 
    \bar{N}_{\rm resc.}\kappa_{4,22\,}\epsilon_2^2,
  \end{equation}
  with a term quadratic in $\epsilon_2$ yet linear in the mean number of rescatterings.
  
  Similarly, one finds
  \begin{equation}\label{v5_vs_eps2,3,sigma}
  v_5 \sim \bar{N}_{\rm resc.}\kappa_{5,5\,}\epsilon_5 + 
    \bar{N}_{\rm resc.}\kappa_{5,23\,}\epsilon_2\epsilon_3,
  \end{equation}
  again with a contribution nonlinear in the initial-state eccentricities and linear in $\bar{N}_{\rm resc.}$.
   
\item Coming to $v_6$, one quickly sees that the approximation~\eqref{approximate_collision-term} will yield a contribution in $\bar{N}_{\rm resc.}\epsilon_3^2$ and one in $\bar{N}_{\rm resc.}\epsilon_2\epsilon_4$.
 However it cannot yield the term in $\epsilon_2^3$ observed in fluid-dynamical studies.

  To recover the latter, one must consider products $f^{(0)}(\bm{i})f^{(1)}(\bm{j})$ in the integrand of the collision term, since $f^{(1)}$ contains a term in $\sigma_{}\epsilon_2^2\cos(4\theta)$---which is reflected in the quadrangular flow~\eqref{v4_vs_eps2,sigma}.
  This will indeed yield a term in $\epsilon_2^3$, but the latter is also proportional to $\sigma^2$:
  \begin{align}\label{v6_vs_eps2,3,sigma}
  v_6 \sim &\ \bar{N}_{\rm resc.}\big(\kappa_{6,6\,}\epsilon_6 + 
    \kappa_{6,33\,}\epsilon_3^2 + 
    \kappa_{6,24\,}\epsilon_2\epsilon_4\big) \cr
    & \ + \bar{N}_{\rm resc.}^2\kappa_{6,222\,}\epsilon_2^3. 
  \end{align}
  At very low $\bar{N}_{\rm resc.}$ the linear and quadratic contributions will dominate, while the cubic term in the ellipticity $\epsilon_2$ will only become meaningful for a larger number of rescatterings.
  
\item Similarly, one can easily convince oneself that the setup consisting of the initial distribution~\eqref{f(t0,x,p)}, evolved with the Boltzmann equation~\eqref{Boltzmann-eq} with collision term~\eqref{collision-integral_2->2} does not generate any contribution to the harmonics $v_{n\geq 7}$ involving only $\epsilon_2$ and $\epsilon_3$ at linear order in $\bar{N}_{\rm resc.}$: 
  the $\epsilon_2^2\epsilon_3$ contribution to $v_7$ is quadratic in $\bar{N}_{\rm resc.}$; 
  $v_8$ will receive a term in $\epsilon_2\epsilon_3^2$ at order $\bar{N}_{\rm resc.}^2$ and a contribution $\epsilon_2^4$ at order $\bar{N}_{\rm resc.}^3$, and so on.
  
\item Eventually, the same reasoning shows that the subleading contribution in $\epsilon_2^3$ to elliptic flow $v_2$, which is observed in fluid dynamical simulations, can also be recovered within our model, at order $\bar{N}_{\rm resc.}^2$.
  More generally, the model predicts a contribution in $\bar{N}_{\rm resc.}^2\epsilon_n^3$ to $v_n$ for any $n$.
\end{itemize} 

Summarizing our findings, we find that our model of a system of self-diffusing particles with an initially asymmetric transverse geometry is able to generate anisotropic flow coefficients $v_n$ with the same scaling dependence on the initial-state eccentricities as within a fluid-dynamical description, as seen from comparing Eqs.~\eqref{v2,v3_vs_eps2,eps3,sigma}--\eqref{v6_vs_eps2,3,sigma} with Eqs.~\eqref{v2,v3_vs_eps2,eps3}--\eqref{v6_vs_eps2,3,6}.
This behavior was already known in the literature~\cite{Borghini:2010hy,Kurkela:2018ygx,Borghini:2011qc}. 

What to our knowledge was never mentioned before regards the scaling of the generated flow harmonics with the rescattering cross section, or equivalently with the mean number of rescatterings per particle.%
\footnote{For completeness, let us note that $\bar{N}_{\rm resc.}$ is roughly the inverse of the Knudsen number in the system.}
Thus, the ``leading contributions'', i.e.\ those stemming from the largest eccentricities $\epsilon_2$ and $\epsilon_3$, to the successive Fourier coefficients $v_n$ scale with different powers of $\bar{N}_{\rm resc.}$.
And, perhaps more interesting, starting with hexagonal flow $v_6$ there can be two or more such leading contributions to $v_n$, which necessarily scale differently with $\bar{N}_{\rm resc.}$:
\begin{equation}\label{v6_vs_eps2,3,Nscat}
v_6 \sim {\cal O}(\bar{N}_{\rm resc.})_{}\epsilon_3^2 + {\cal O}(\bar{N}_{\rm resc.}^2)_{}\epsilon_2^3,
\end{equation}
where the terms in $\epsilon_2\epsilon_4$ or $\epsilon_6$ are assumed to be smaller.
That is, the development of the contribution to $v_6$ from the ellipticity $\epsilon_2$ necessitates more rescatterings than that of the triangularity $\epsilon_3$.
Similarly, one easily finds $v_8 \sim {\cal O}(\bar{N}_{\rm resc.}^2)_{}\epsilon_2\epsilon_3^2 + {\cal O}(\bar{N}_{\rm resc.}^3)_{}\epsilon_2^4$, again assuming that the contributions to $v_8$ involving eccentricities $\epsilon_p$ with $p\geq 4$ are small.

\section{Discussion}
\label{s:discussion}

In this last section, we address two issues: 
first, are our results on the scalings with the average number of rescatterings $\bar{N}_{\rm resc.}$, in particular that of Eq.~\eqref{v6_vs_eps2,3,Nscat}, robust against changes of the setup which we considered? 
And second, is there a possibility to evidence these behaviors in experimental data?

If anisotropic flow is not present initially, but generated by rescatterings, then the corresponding harmonics will depend on the initial-state eccentricities $\epsilon_n$ and on $\bar{N}_{\rm resc.}$. 
Regarding the linear response of $v_n$ to $\epsilon_n$, we cannot think of a plausible scenario in which it would not already be generated at linear order in $\bar{N}_{\rm resc.}$, i.e.\ $v_n = {\cal O}(\bar{N}_{\rm resc.})_{}\epsilon_n$, generalizing Eq.~\eqref{v2,v3_vs_eps2,eps3,sigma}.
Less straightforward are the nonlinear response behaviors, which we now discuss at length. 

The emergence of scalings $v_{n+p} \propto \epsilon_n\epsilon_p$ at linear order in $\bar{N}_{\rm resc.}$, as in Eqs.~\eqref{v4_vs_eps2,sigma}--\eqref{v5_vs_eps2,3,sigma}, is also straightforward in a description in which the collision term ${\cal C}[f]$ is at least quadratic in the single-particle distribution $f$, which is a natural feature in a picture in which the momentum anisotropies are generated by rescatterings of at least two partners. 

The less trivial scaling behavior is that of Eq.~\eqref{v6_vs_eps2,3,Nscat}, in particular the ${\cal O}(\bar{N}_{\rm resc.}^2)$ dependence of the term in $\epsilon_2^3$.
To investigate whether it is an artifact of our model or more general, we note that only two types of modifications are possible as long as one remains in a kinetic-theoretical framework with particle scatterings: 
changes in the initial-state distribution $f(t_0,{\bf x},{\bf p})$, which entirely determines the free-streaming solution $f^{(0)}(t,{\bf x},{\bf p})$, or of the collision term ${\cal C}[f]$.
In both cases, we want to see how a term in $\epsilon_2^3$ might arise at first order in $\bar{N}_{\rm resc.}$ or somewhat equivalently in the interaction cross section $\sigma$, thereby invalidating Eq.~\eqref{v6_vs_eps2,3,Nscat}.

Changing the initial-state distribution so as to spoil Eq.~\eqref{v6_vs_eps2,3,Nscat} is mathematically feasible, by assuming that $F({\bf p})$ contains a term in $\epsilon_2$. 
This would however mean that the initial momentum distribution already knows about the global geometry of the collision zone, which is problematic from the physics point of view.
In turn, if the isotropic term $G(r)$ contains a term in $\epsilon_2$, then the latter will multiply both terms of Eq.~\eqref{v6_vs_eps2,3,Nscat}, which is also not what we wish. 

Accordingly, the only viable modifications to be considered are changes of the collision term ${\cal C}[f]$.
Sticking to the general structure of a collision integral involving $f$,\footnote{We could not come up with a setup involving a non-factorized two-particle phase space distribution $f_2(\bm{i},\bm{j})$ in the integrand, instead of a product of single-particle distributions, that would {\it generate\/} both contributions to $v_6$ at first order in $\bar{N}_{\rm resc.}$.} one quickly sees that the generalization of Boltzmann's ansatz~\eqref{collision-integral_2->2} necessary to obtain a term $\epsilon_2^3$ at first order in $\bar{N}_{\rm resc.}$ is to include a contribution of (at least) cubic order in $f$ in the integrand. 
Two kinds of physical causes justify such contributions.
On the one hand, one may include rescatterings with at least three particles in the initial state, in particular three-to-two scatterings, as can be found e.g.\ in Eq.~(12) of Ref.~\cite{Xu:2004mz}.
Here, one should note that including two-to-three scatterings only would not help.
In addition, three-to-two rescatterings will in fact generate the desired term in $\epsilon_2^3$ at first order in the corresponding cross section $\sigma_{3\to 2}$:
whether the latter yields the leading contribution to $\bar{N}_{\rm resc.}$, so that the term is indeed of order ${\cal O}(\bar{N}_{\rm resc.})_{}\epsilon_2^3$, or else $\bar{N}_{\rm resc.}$ is rather dominated by two-to-two rescatterings becomes a partly model-dependent issue. 

On the other hand, even restricting oneself to elastic two-to-two rescatterings, one may still consider the generalization of the integrand accounting for Bose--Einstein enhancement or Pauli blocking, i.e.\ with factors of the form $f(\bm{i})f(\bm{j})[1\pm f(\bm{k})][1\pm f(\bm{l})]$ in lieu of $f(\bm{i})f(\bm{j})$.

The inclusion of three-to-two scatterings and/or quantum-mechanical phase-space occupancy effects seems at first face to be relevant only if the initial state is that of a dense system. 
Naturally, when the latter expands, it becomes more dilute, and the terms beyond quadratic order in $f$ in the collision integral become less important.
Nevertheless, it is not clear to us whether a system created in high-energy collisions could be in a regime such that the emitted particles rescatter only very little, while at the same time being initially dense and interacting enough to lead to a breakdown of Eq.~\eqref{v6_vs_eps2,3,Nscat}. 
``Small systems'' are the most likely candidates for such a departure, provided the initial density is big. 
In any case, the relative importance of more-than-quadratic terms and their influence on the scaling behavior could be tested in numerical simulations with transport codes in which they can be switched on or off at will, like $2\leftrightarrow 3$ scatterings in BAMPS~\cite{Xu:2004mz} or quantum effects in other codes~\cite{Zhang:2017esm}.

Let us now discuss where in experimental data the scaling behaviors~\eqref{v2,v3_vs_eps2,eps3,sigma}--\eqref{v6_vs_eps2,3,Nscat} could possibly be at play and measurable. 

Surprising though it may seem, let us first deal with larger systems, in which fluid dynamics is routinely applied to describe the evolution. 
On the one hand, the single-collision regime might be applicable to particles in given regions of phase space, e.g.\ at high transverse momentum~\cite{Romatschke:2018wgi}, or to specific particle types, like bottomonia---for which the picture is rather a negative one: a collision means destruction. 
On the other hand, Eqs.~\eqref{v2,v3_vs_eps2,eps3,sigma}--\eqref{v6_vs_eps2,3,Nscat} may also be relevant in phenomenological analyses of the bulk of particles. 
More accurately, these behaviors play a role in the pre-hydrodynamized stage, which in modern hybrid descriptions is often modeled by a transport cascade~\cite{Liu:2015nwa}. 
Indeed, this short kinetic period, which only involves rather few rescatterings, will transform ``pre-early transport eccentricities'', taken from a model for initial conditions, into some early anisotropic flow, which becomes part of the initial condition for the fluid-dynamical stage of the hybrid description. 
Following Eq.~\eqref{v6_vs_eps2,3,Nscat}, the early generated $v_6$ will suffer from a deficit in second-order eccentricity $\epsilon_2$, which will be propagated by the subsequent evolution until the final state.  
This could then affect attempts at evidencing the scaling~\eqref{v6_vs_eps2,3,6} and at interpreting it within a purely fluid-dynamical framework, since the coefficients ${\cal K}_{6,33}$ and ${\cal K}_{6,222}$ will contain a pre-hydrodynamization component, whose relative size possibly varies across centralities. 
This possibility is a further incentive to investigate the scaling behaviors~\eqref{v2,v3_vs_eps2,eps3,sigma}--\eqref{v6_vs_eps2,3,Nscat} in transport models.

Eventually, the natural place where Eqs.~\eqref{v2,v3_vs_eps2,eps3,sigma}--\eqref{v6_vs_eps2,3,Nscat} are to be looked for is in small systems, in which the applicability of fluid dynamics is most questionable. 
The biggest issue is of course that the anisotropic flow coefficients in such systems are small. 
We believe that the more trivial scaling behaviors~\eqref{v2,v3_vs_eps2,eps3,sigma}--\eqref{v5_vs_eps2,3,sigma} should be rather ``easily'' observable. 
Note in particular that the unknown mean number of rescatterings cancels in the ratio of two different harmonics $v_2$--$v_5$, so that one can separate the influences of eccentricities and $\bar{N}_{\rm resc.}$, where one can expect that the latter should scale like the cubic root of the charged particle multiplicity $(\dd N^{\rm ch.\!}/\dd\eta)^{1/3}$.
In turn, we are aware that in small systems $v_6$ will be at the border of what is measurable with reasonable uncertainties, so that whether measurements allowing to test Eq.~\eqref{v6_vs_eps2,3,Nscat} are feasible is not warranted.
Nevertheless, we think that confirming the scaling behavior~\eqref{v6_vs_eps2,3,Nscat} would yield further confidence in the determined value of $\bar{N}_{\rm resc.}$, at the same time evidencing a nice instance of nonlinearity.
Conversely, as we have already discussed above, departure from that behavior might hint at a dense initial state, possibly saturated, which is certainly not an uninteresting result and is worth investigating.

\begin{acknowledgments}
  We thank Kai Gallmeister and Carsten Greiner for precisions regarding the collision kernel in BAMPS. 
  We acknowledge support by the Deutsche Forschungs\-gemeinschaft (DFG) through the grant CRC-TR 211 ``Strong-interaction matter under extreme conditions''.
\end{acknowledgments}

\end{document}